\documentclass[pra,twocolumn,showpacs,preprintnumbers]{revtex4}
\usepackage{amsmath}
\usepackage{graphicx}
\usepackage{mathrsfs}
\usepackage{amssymb}
\usepackage{amsfonts}
\usepackage{amsmath}
\usepackage[colorlinks,citecolor=blue,linkcolor=red,hyperindex,CJKbookmarks,dvipdfm]{hyperref}

\begin{document}

\title{Adiabatic Mach-Zehnder interferometer via an array of trapped ions}
\author{Y. M. Hu$^{1,2,3}$}
\altaffiliation{Electronic address: huyanmin2003@163.com}
\author{M. Feng$^{1}$}
\altaffiliation{Electronic address: mangfeng@wipm.ac.cn}
\author{C. Lee$^{3}$}
\altaffiliation{Electronic address: chleecn@gmail.com}
\affiliation{$^{1}$State Key Laboratory of Magnetic Resonance and Atomic and Molecular
Physics, Wuhan Institute of Physics and Mathematics, Chinese Academy of
Sciences, and Wuhan National Laboratory for Optoelectronics, Wuhan 430071,
China}
\affiliation{$^{2}$Graduate School of the Chinese Academy of Sciences, Beijing
100049,China}
\affiliation{$^{3}$State Key Laboratory of Optoelectronic Materials and Technologies,
School of Physics and Engineering, Sun Yat-Sen University, Guangzhou 510275,
China}
\date{\today}

\begin{abstract}
We explore the possibility of implementing a Heisenberg-limited Mach-Zehnder
interferometry via an array of trapped ions, which obey a quantum Ising
model within a transverse field. Based upon adiabatic processes of
increasing the Ising interaction and then decreasing the transverse field,
we demonstrate a perfect transition from paramagnetism to ferromagnetic
states, which can be used as the beam splitter for the multi-ion
Mach-Zehnder interferometry. The achieved NOON state of the ions enables the
Heisenberg-limited interferometry. Using currently available techniques for
ultracold ions, we discuss the experimental feasibility of our scheme with
global operations.
\end{abstract}

\maketitle

Since the foundation of quantum theory, as a natural result of superposition
principle, quantum interference has attracted continuous attentions in both
theoretical and experimental studies. It has been widely used to implement
high-precision measurement, quantum information processing and so on. A
well-known scheme for performing quantum interferometry is the Mach-Zehnder
(MZ) interferometry, which has a beam splitter for splitting the input
states and another beam splitter for recombining the output states. Up to
now, quantum MZ interferometry has been accomplished via photons~\cite%
{photon}, electrons~\cite{electron}, superconducting flux qubits~\cite%
{superconducting} and trapped ions~\cite{ion1,ion2} etc.

Beyond the conventional quantum interferometry via unentangled states, it
has been demonstrated that the measurement precision can be enhanced from
the standard quantum limit (or the shot noise limit) to the Heisenberg limit
by using multipartite entangled states~\cite{Giovannetti}. An excellent
candidate is the NOON state $\left(\left\vert N\right\rangle _{a}\left\vert
0\right\rangle _{b} + \left\vert 0\right\rangle_{a}\left\vert
N\right\rangle_{b}\right)/\sqrt{2}$, which is an equal-probability
superposition of all $N$ particles in one of two paths denoted by $a$ and $b$%
. The entangled ions for high-precision metrology have been proposed
theoretically~\cite{Wineland} and demonstrated experimentally~\cite%
{Wineland1,Leibfried}. However, these schemes are subject to limited numbers
of ions or the requirement for individual addressing. For an ensemble of
thousands of neutral atoms, the possibility of performing a
Heisenberg-limited MZ interferometry has been demonstrated via a quantized
Bose-Josephson junction~\cite{Lee2006}.

In this article, based upon the adiabatic processes and global
operations on an array of ultracold ions, we present a realizable scheme for
performing a Heisenberg-limited MZ interferometry. The ion array is
described by a quantum Ising model of ferromagnetic (FM) interaction $J$ and
transverse field $B$, which has been used to simulate quantum magnetism~\cite%
{Mg,monroe1,monroe2} and quantum phase transition (QPT)~\cite%
{monroe3,monroe4,monroe5} by virtue of spin-dependent optical dipole forces~%
\cite{porras}. In our scheme, if the system starts from a paramagnetic state
dominated by $B$, we adiabatically increase $J$ and then decrease $B$, and
vice versa if the system starts from a FM state dominated by $J$. In
contrast to tuning either $J$ or $B$ in previous schemes~\cite{Mg, monroe5},
the adiabatic processes in our scheme perfectly connects the two limits
solely controlled by one of $B$ and $J$. Therefore, theoretically, a pure
NOON state can be prepared adiabatically from a SU(2) coherent state, which
is the ground state for the system completely dominated by $B$. Our scheme
requests only global operations, which is less challenging experimentally
than individual addressing of the ions. By using the adiabatic process
between paramagnetic and FM states as beam splitters, we are able to
accomplish a Heisenberg-limited MZ interferometry via the NOON state of an
array of trapped ions.

We consider N ultracold ions confined in a linear Paul trap. Because only
two ionic hyperfine states are used for implementing the MZ interferometry,
each ion can be regarded as a spin-1/2 particle of two spin states $%
\left\vert \downarrow \right\rangle $ and $\left\vert \uparrow \right\rangle
$. By globally addressing all the ions with particular lasers, the system
obeys a transverse-field quantum Ising Hamiltonian~\cite{monroe3,monroe4,
monroe5,porras},
\begin{equation}
H=-\sum_{i<j}J_{ij}\sigma _{z}^{i}\sigma _{z}^{j}-B\sum_{i}\sigma _{x}^{i},
\end{equation}%
where $\sigma _{x,z}^{i}$ are Pauli operators for the $i$-th ion, $B$ is the
transverse field, and $J_{ij}=J/\left\vert i-j\right\vert ^{3}$ is the
effective Ising interaction between ions $i$ and $j$ with $J\geq 0$ denoting
the nearest-neighboring interaction. For convenience, we consider the
dimensionless model in units of $\hbar =1$ before the discussions of
experimental possibility. This model has been experimentally realized by
using either longitudinal~\cite{porras,Mg} or transverse modes \cite%
{monroe1, monroe2, monroe3, monroe4, monroe5} of the ions. Obviously, there
are two extreme cases: $J=0$ and $B=0$. If $J=0$, the ground state is a
paramagnetic state of all spins aligned with the magnetic field, i.e., $%
\left\vert \rightarrow \rightarrow \cdots \rightarrow \right\rangle $ for $%
B>0$ or $\left\vert \leftarrow \leftarrow \cdots \leftarrow \right\rangle $
for $B<0$. This ground state is a spin coherent state. If $B=0$, there are
two degenerate ground states of all spins in either the spin-down state $%
\left\vert \downarrow \downarrow \cdots \downarrow \right\rangle $ or the
spin-up state $\left\vert \uparrow \uparrow \cdots \uparrow \right\rangle $.
The equal-probability superposition of these two degenerate ground states is
a NOON state.

\begin{figure}[tbh]
\begin{center}
\includegraphics[width=1.0 \columnwidth]{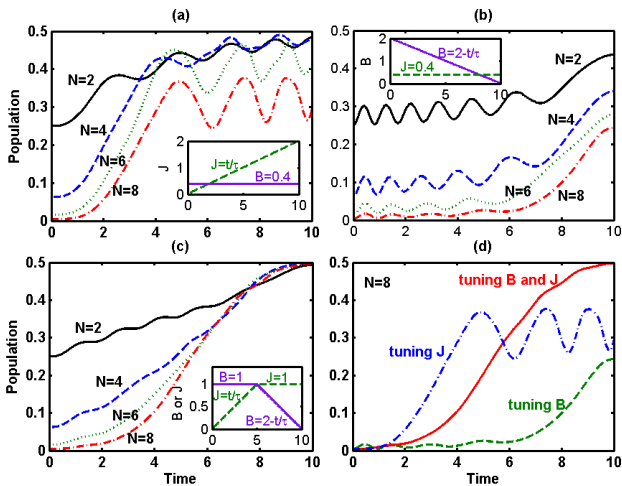}
\end{center}
\par
\label{Fig.1}
\caption{(Color online) Population evolutions of FM states for systems of
different $N$, in which the time $t\in [0, 2\protect\tau]$ with $\protect%
\tau = 5$. (a) $J = t/\protect\tau $ and $B=0.4$. (b) $J=0.4$ and $B=2-t/
\protect\tau$. (c) $J=t/\protect\tau$ and $B=1$ if $t\in [0, \protect\tau]$,
and $J=1$ and $B=2-t/\protect\tau$ if $t\in [\protect\tau ,2\protect\tau]$.
(d) Population evolutions for $N=8$ under different sweeping schemes. In
(a), (b) and (c), the insets stand for the temporal modulations of $B$
and/or $J$.}
\end{figure}

To prepare the NOON state for the multi-ion MZ interferometry under global
operations, one may use a quantum adiabatic evolution connecting the initial
spin coherent state and the NOON state. Obviously, the initial state is the
ground state for the system of $J=0$ and $B>0$ and the desired NOON state is
the ground state for the system of $J>0$ and $B=0$. Two sweeping schemes
have been suggested for preparing the NOON state with high fidelities. In
the first scheme, $J$ is adiabatically increased from 0 to $B/J\ll 1$ (with
constant $B$)~\cite{Mg}. In the second scheme, $B$ is adiabatically
decreased from $B/J\gg 1$ to 0 (with constant $J$)~\cite{monroe5}. Starting
from $\left\vert \rightarrow \rightarrow \cdots \rightarrow \right\rangle $,
we show the dynamical populations in the two FM states $\left\vert
\downarrow \downarrow \cdots \downarrow \right\rangle $ and $\left\vert
\uparrow \uparrow \cdots \uparrow \right\rangle $, see Fig.~1(a) and (b).
The two FM states always have the same populations due to the absence of
longitudinal fields and the fidelity to the desired NOON state is associated
with the sum of these two populations. The final populations depend on the
sweeping rate and the system size. But even if the time-evolution is
perfectly adiabatic, for a finite-size system, the final fidelity to the
NOON state can only approach unity but not exactly unity. This is because
that the final ground state of the first scheme~\cite{Mg} is not exactly the
FM states and the initial state is not the ground state for the second
scheme~\cite{monroe5}.

To improve the fidelity to the desired NOON state, different from the
single-step schemes of sweeping $B$ or $J$, we adopt in the present work a
two-step scheme exactly connecting two extreme cases, i.e., $(J=0,~B>0)$ and
$(J>0,~B=0)$, in which $J$ and $B$ are alternately changed in one of the two
steps. Thus, the ground state for the initial Hamiltonian $%
H_{i}=-B\sum_{i}\sigma _{x}^{i}$ is exactly the spin coherent state along $%
x- $axis, and the NOON state is exactly a ground state for the final
Hamiltonian $H_{f}=-\sum_{i<j}J_{ij}\sigma _{z}^{i}\sigma _{z}^{j}$. In the
first step, the ratio $J/B$ increases from 0 to 1 with $B$ remaining
unchanged. While in the second step, the ratio $B/J$ decreases from 1 to 0
with $J$ remaining unchanged. In our calculations, the initial state is
chosen as $\left\vert \rightarrow \rightarrow \cdots \rightarrow
\right\rangle $, which can be easily prepared. The system first evolves with
$B=1$ and $J=t/\tau $ from $t=0$ to $t=\tau $ and then evolves with $J=1$
and $B=2-t/\tau $ from $t=\tau $ to $t=2\tau $. Therefore, for a
sufficiently large $\tau $, the system will adiabatically evolve into the
desired NOON state $\left\vert \Psi \right\rangle =(\left\vert \downarrow
\downarrow \cdots \downarrow \right\rangle +\left\vert \uparrow \uparrow
\cdots \uparrow \right\rangle )/\sqrt{2}$ with very high fidelities, see
Fig.~1(c) and Fig.~2. Besides, Fig.~1(d) shows that the NOON state achieved
by our two-step scheme is much better than the ones achieved by the
single-step schemes.

\begin{figure}[tbh]
\begin{center}
\includegraphics[width=1.0 \columnwidth]{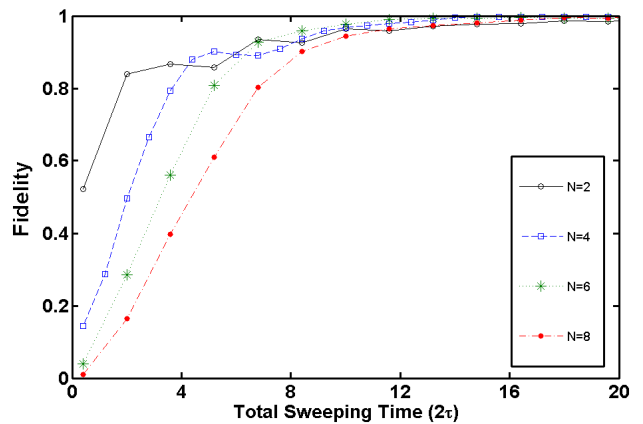}
\end{center}
\caption{(Color online) Fidelity to the NOON state versus the total sweeping
time $2\protect\tau $ for systems of different $N$. Experimentally the
fidelity is relevant to the ferromagnetic order parameter, which could be
obtained by measuring the populations of the spin-up and -down states with spin-dependent fluorescence \protect\cite{monroe5}.}
\label{Fig.2}
\end{figure}

On the other hand, at the expense of a longer total sweeping time ($2\tau $%
), the single-step scheme of tuning $B$ or $J$ may also achieve a desired
fidelity. However, this is not experimentally practical for multi-ion
systems restricted by the limited coherent time. Therefore, our two-step
scheme of a shorter total sweeping time seems a better choice for preparing
the desired NOON state.

Utilizing the prepared NOON state, we may achieve a Heisenberg-limited MZ
interferometry by using $\left\vert \uparrow \uparrow \cdots \uparrow
\right\rangle $ and $\left\vert \downarrow \downarrow \cdots \downarrow
\right\rangle $ as two paths and by global operations for all the ions. The
schematic diagram for our adiabatic MZ interferometry is shown in Fig. 3, in
which a free evolution is sandwiched by two beam splitters achieved by
adiabatic processes connecting paramagnetic and FM states. The first beam
splitter (BS1) is the adiabatic preparation of the NOON state $\left\vert
\Psi \right\rangle $ discussed above. Then the state $\left\vert \Psi
\right\rangle $ evolves into $\left\vert \Psi ^{^{\prime }}\right\rangle
=(e^{-iN\phi /2}\left\vert \uparrow \uparrow \cdots \uparrow \right\rangle
+e^{iN\phi /2}\left\vert \downarrow \downarrow \cdots \downarrow
\right\rangle )/\sqrt{2}$ in the following free evolution of time duration $%
T $ under the government of $H_{0}=\omega _{0}\sum_{i}\sigma _{z}^{i}$,
where $\omega _{0}$ is the transition frequency between $\left\vert \uparrow
\right\rangle $ and $\left\vert \downarrow \right\rangle $ and the
accumulated phase $\phi =\omega _{0}T$. Lastly, the second beam splitter
(BS2) is accomplished by the inverse process of the BS1.

\begin{figure}[tbh]
\begin{center}
\includegraphics[width=1.0 \columnwidth]{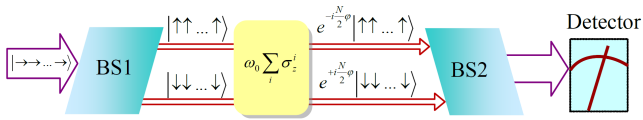}
\end{center}
\caption{(Color online) The schematic diagram for the adiabatic MZ
interferometry via trapped ions. Here, BS1 prepares the NOON state via
adiabatic processes connecting paramagnetic and FM states, then two paths
accumulate a relative phase in a following free evolution, and BS2
recombines the two paths for interference.}
\label{Fig.3}
\end{figure}

To extract the accumulated relative phase between the two paths, one has to
transform the phase information into the amplitude information of the final
state itself or the expectation information of a particular observable for
the final state. For systems of few trapped ions, this could be done by a
controlled-NOT gate with the first ion being the control qubit and the rest
being the target ones followed by a Hadamard operation on the first ion~\cite%
{Wineland}. However, the individual addressing used in this procedure is
very difficult to be accomplished in systems of large numbers of particles.
In our scheme with full global operations, we may accomplish this procedure
in another adiabatic process, which is the inverse process of the BS1, i.e.,
by first slowly increasing $B/J$ from 0 to 1 (with $J$ fixed) and then
slowly decreasing $J/B$ from 1 to 0 (with $B$ fixed). As a result, for
extracting the phase shift from our multi-ion MZ interferometry, we have to
measure populations of the ground state $\left\vert 0\right\rangle $ and the
first excited state $\left\vert 1\right\rangle $ of the Hamiltonian $H_{i}$~%
\cite{explain}. Under a global Hadamard operation, $\left\vert
0\right\rangle $ turns to be spin-up with all the ions, and $\left\vert
1\right\rangle $ becomes an entangled state with one spin-down and others
spin-up. By applying a probe laser for coupling $|\downarrow \rangle $ to an
excited level, it is possible to distinguish $\left\vert 1\right\rangle $
from $\left\vert 0\right\rangle $ by detecting spin-dependent fluorescence
signals~\cite{monroe5}. Therefore, we could obtain the population of the
first excited state,%
\begin{equation}
P_{1}=\sin ^{2}(N\phi /2),
\end{equation}
and the corresponding phase sensitivity is
\begin{equation}
\Delta \phi =\Delta \widehat{P}_{1}/(\partial \left\langle \widehat{P}_{1}%
\right\rangle /\partial \phi )=1/N,
\end{equation}
with the population operator $\widehat{P}_{1}=\left\vert 1\right\rangle
\left\langle 1\right\vert$ and its variance $\Delta \widehat{P}_{1}$. This is the Heisenberg limit, where the
relative phase $\phi $ is measured more precisely than the disentanglement
case by $\sqrt{N}$ times. Since $\phi =\omega _{0}T$ and $\omega _{0}$
depends on the atomic configuration and the local magnetic field, our
Heisenberg-limited MZ interferometry is of practical applications in
understanding the atomic configuration and measuring the local magnetic
field.

Experimentally, the transverse-field quantum Ising model could be realized
by an array of ultracold ions which are globally illuminated with
off-resonant lasers and resonant Raman beams~\cite%
{monroe1,monroe2,monroe3,monroe4,monroe5,porras}. Because of the
off-resonant lasers, the inter-ion Ising interactions are induced by the
spin-dependent forces~\cite{LD} with the assistance of phonon modes. The
effective transverse fields are generated by the resonant Raman beams, which
resonantly couple the two spin states. In the rotating frame, the ionic
array under the Lamb-Dicke limit is equivalent to a quantum Ising system
within a transverse magnetic field. With the control of the frequencies and
intensities of the off-resonant lasers, the sign and strength of the Ising
interaction can be dynamically adjusted~\cite{monroe4}. By tuning the
strength of the resonant Raman beams, we can dynamically adjust the strength
of the transverse field \cite{monroe2,monroe3,monroe5}.

For an array of $^{171}$Yb$^{+}$ ions, two clock states $^{2}\text{S}%
_{1/2}\left\vert F=0,m_{F}=0\right\rangle$ and $^{2}\text{S}_{1/2}\left\vert
F=1,m_{F}=0\right\rangle$ are employed as the two spin states $%
\left\vert\downarrow\right\rangle$ and $\left\vert\uparrow\right\rangle$ for
a spin-1/2 particle, respectively. Based upon an array of ultracold $^{171}$%
Yb$^{+}$ ions in a linear trap, the QPT in the transverse-field quantum
Ising model has been simulated~\cite{monroe5}, in which both $B$ and $J$ may
be larger than the order of kHz. Therefore, we assume the maximum value for $%
B$ and $J$ is $B_0=J_0=50$ kHz for our two-step sweeping scheme shown in the
inset of Fig.~1(c). To compare with the dimensionless model, we transform
the original Hamiltonian $H$ into $H/J_0$, which means that the energy is in
units of $J_0$ and the time is in units of $1/J_0$. In the first step, we
fix $B=50$ kHz and slowly vary $J$ according to $J=J_{0}\times $t/$\tau$
from $t=0$ to $t = \tau$, where $\tau = 100 \mu$s. Then we fix $J=50$ kHz
and slowly sweep $B$ following $B=B_{0}\times(2-t/\tau)$ from $t=\tau$ to $%
t=2\tau$. Besides, to maintain the coherence of the NOON state, the free
evolution time $T$ should be much shorter than $\tau$. Therefore the
required operations from BS1 to BS2 in our proposed MZ interferometry take
about 400 $\mu$s, which is shorter than the gating time (longer than 500 $%
\mu $s) in Ref.~\cite{monroe5} and should be feasible with currently
available techniques for finite spin size.

Our scheme could also be realized by using other types of ions. For an array
of $^{40}$Ca$^{+}$ ions in a linear trap, $|\downarrow\rangle$ and $%
|\uparrow\rangle$ are encoded by two hyperfine states $|S_{1/2},~m_{S}=-1/2%
\rangle$ and $|S_{1/2},m_{S}=1/2\rangle$ [or $|S_{1/2},~m_{S}=-1/2\rangle$
and the metastable state $|D_{5/2},~m_{S}=-1/2\rangle$], respectively. It
has demonstrated the entanglement of fourteen $^{40}$ Ca$^{+}$ ions in a
linear trap~\cite{blatt}. More recently, the universal digital quantum
simulation based upon a transverse-field quantum Ising model has been
implemented by using $^{40}$Ca$^{+}$~\cite{blatt2}. For a system of fourteen
$^{40}$Ca$^{+}$ ions, our interferometry scheme could enhance the
measurement precision by nearly four times in comparison to the standard
quantum limit.

\begin{figure}[tbh]
\begin{center}
\includegraphics[width=1.0 \columnwidth]{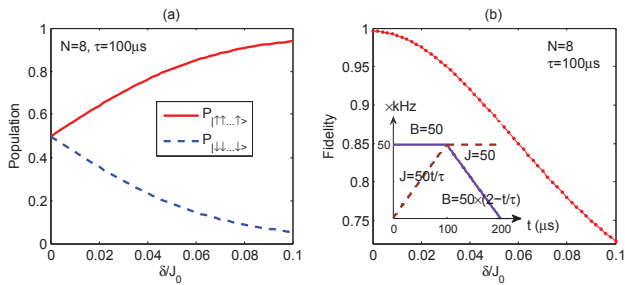}
\end{center}
\caption{(Color online) The undesirable impact from the deviation in the
process of BS1 for eight ions by sweeping time $t=2\protect\tau =200\protect%
\mu s$. (a) Populations in $\left\vert \uparrow \uparrow \cdots \uparrow
\right\rangle $ and $\left\vert \downarrow \downarrow \cdots \downarrow
\right\rangle $ with respect to $\protect\delta /J_{0}$, and (b) the
fidelity of the prepared state to the ideal NOON state with respect to $%
\protect\delta /J_{0}$ with the variation of B and J in the inset during the
adiabatic evolution.}
\label{Fig.4}
\end{figure}

Besides unpredictable imperfection in operations, the main errors in
realistic experiments are caused by spontaneous emission and
intensity/frequency fluctuations of the Raman beams, which lead to
decoherence and suppress the populations of FM states~\cite{monroe5}. The
spontaneous emission in our scheme can be minimized by enlarging the Raman
detuning from the excited state, provided that we increase the laser power
accordingly to maintain the original Ising interaction and the transverse
field strength. The intensity fluctuations of the Raman beams only yield AC
Stark shift with $2\%$ rms error~\cite{monroe1}, which is not serious in our
scheme. Alternatively, we may employ $^{40}$Ca$^{+}$ ions irradiated by a
729 nm laser to couple the ground state $|S_{1/2},~m_{S}=-1/2\rangle $ with
the metastable state $|D_{5/2},~m_{S}=-1/2\rangle $, which completely remove
the necessity of Raman beams in manipulation~\cite{riebe}. For the
operational imperfection, we simply consider a typical one in the
non-resonant carrier transition when we tune the parameter $B$, which yields
unwanted longitudinal bias field $\delta \sum_{i}\sigma _{z}^{i}$ and may
cause a deviation from the expected NOON state. This imperfection is shown in Fig. 4, and it could be suppressed by
carefully adjusting the direction and the polarization of the laser beams
and/or by using additional pulses~\cite{Mg} to compensate the deviation.

In our scheme, the two beam splitters form a loop-like adiabatic operation,
which may generate a geometric phase~\cite{geometric}. Fortunately, the
generated geometric phase is a constant relevant to the loop area, but do
not depend on the free evolution time $T$ (i.e. the accumulated phase $%
\omega _{0}T$)~\cite{Erik}. Therefore, by measuring the relative phase for
different free evolution time, we may eliminate the influence of the
geometric phase by reducing a common shift from measurement results and the
measurement precision of $\omega _{0}$ is thereby independent from the
geometric phase accumulated in the adiabatic operation.

In conclusion, we have proposed a simple and practical scheme to carry out a
Heisenberg-limited MZ interferometer with an array of trapped ions, which
obey a transverse-field quantum Ising model. The multi-ion MZ interferometry
is implemented by only global operations, which favors scalability. However,
more challenges would appear if more ions are involved, such as less
homogeneous laser irradiation on the ions, smaller energy gap in the Ising
model and weaker laser-ion coupling. Nevertheless, for finite ions, the
global operation is less difficult experimentally than individual
addressing. Provided the nearly perfect adiabaticity, our scheme includes a
QPT in each beam splitter. By suppressing unpredictable errors in realistic
experiments and elaborately modifying the results, our scheme for quantum
metrology is in principle able to reach the ultimate precision limit beyond
the standard quantum limit \cite{Wineland2}.

This work is supported by the NBRPC under Grants No. 2012CB821305 and No.
2012CB922102, the NNSFC under Grants No. 10974225, No. 11075223 and No.
11004226, the NCETPC under Grant No. NCET-10-0850 and the Fundamental
Research Funds for Central Universities of China.


\begin{thebibliography}{99}
\bibitem{photon} J. G. Rarity, P. R. Tapster, E. Jakeman, T. Larchuk, R. A.
Campos, M. C. Teich, and B. E. A. Saleh, Phys. Rev. Lett. \textbf{65}, 1348
(1990).

\bibitem{electron} Y. Ji, Y. Chung, D. Sprinzak, M. Heiblum, D. Mahalu, and
H. Shtrikman, Nature (London) \textbf{422}, 415 (2003).

\bibitem{superconducting} W. D. Oliver, Y. Yu, J. C. Lee, K. K. Berggren, L.
S. Levitov, and T. P. Orlando, Science \textbf{310}, 1653 (2005).

\bibitem{ion1} R. Huesmann, Ch. Balzer, Ph. Courteille, W. Neuhauser, and P.
E. Toschek, Phys. Rev. Lett. \textbf{82}, 1611 (1999).

\bibitem{ion2} D. Leibfried, B. DeMarco, V. Meyer, M. Rowe, A. Ben-Kish, J.
Britton, W. M. Itano, B. Jelenkovi\'{c} C. Langer, T. Rosenband, and D. J.
Wineland, Phys. Rev. Lett.\textbf{\ 89}, 247901 (2002).

\bibitem{Giovannetti} V. Giovannetti, S. Lloyd, and L. Maccone, Science
\textbf{306}, 1330 (2004); Phys. Rev. Lett. \textbf{96}, 010401 (2006).

\bibitem{Wineland} J. J. Bollinger, W. M. Itano, D. J. Wineland, and D. J.
Heinzen, Phys. Rev. A \textbf{54}, R4649 (1996); S. Boixo, A. Datta, M. J.
Davis, A. Shaji, A. B. Tacla, and C. M. Caves, Phys. Rev. A \textbf{80},
032103 (2009).

\bibitem{Wineland1} C. A. Sackett, D. Kielpinski, B. E. King, C. Langer, V.
Meyer, C. J. Myatt, M. Rowe, Q. A. Turchette, W. M. Itano, D. J. Wineland,
and C. Monroe, Nature (London) \textbf{404}, 256 (2000).

\bibitem{Leibfried} D. Leibfried, M. D. Barrett, T. Schaetz, J. Britton, J.
Chiaverini, W. M. Itano, J. D. Jost, C. Langer, and D. J. Wineland, Science
\textbf{304}, 1476 (2004).

\bibitem{Lee2006} C. Lee, Phys. Rev. Lett.\textbf{\ 97}, 150402 (2006); C.
Lee, J. Huang, H. Deng, H. Dai, J. Xu, Front. Phys. \textbf{7}, 109 (2012).

\bibitem{Mg} A. Friedenauer, H. Schmitz, J. T. Glueckert, D. Porras, and T.
Schaetz, Nature Phys. \textbf{4}, 757 (2008).

\bibitem{monroe1} K. Kim, M.-S. Chang, R. Islam, S. Korenblit, L.-M. Duan,
and C. Monroe, Phys. Rev. Lett. \textbf{103}, 120502 (2009).

\bibitem{monroe2} K. Kim, M.-S. Chang, S. Korenblit, R. Islam, E. E.
Edwards, J. K. Freericks, G.-D. Lin, L.-M. Duan, and C. Monroe, Nature
(London) \textbf{465}, 590 (2010).

\bibitem{monroe3} E. E. Edwards, S. Korenblit, K. Kim, R. Islam, M.-S.
Chang, J. K. Freericks, G.-D. Lin, L.-M. Duan, and C. Monroe, Phys. Rev. B
\textbf{82}, 060412(R) (2010).

\bibitem{monroe4} G.-D. Lin, C. Monroe, and L.-M. Duan, Phys. Rev. Lett.
\textbf{106}, 230402 (2011).

\bibitem{monroe5} R. Islam, E.E. Edwards, K. Kim, S. Korenblit, C. Noh, H.
Carmichael, G.-D. Lin, L.-M. Duan, C.-C. Joseph Wang, J. K. Freericks, and
C. Monroe, Nat. Commun. \textbf{2}, 377 (2011).

\bibitem{porras} D. Porras and J. I. Cirac, Phys. Rev. Lett. \textbf{92},
207901 (2004); X.-L. Deng, D. Porras, and J. I. Cirac, Phys. Rev. A \textbf{%
72}, 063407 (2005).


\bibitem{explain} For $N$ ions, the ground state $|0\rangle=[(|\uparrow
\rangle+|\downarrow\rangle)/\sqrt{2}]^{\otimes N}$ and the first excited
state $|1\rangle$ consists of $N$ degenerate sub-states. For a system of $%
N=3$, its three degenerate sub-states are $|e_{1}\rangle=(-|\uparrow%
\uparrow \uparrow\rangle+|\uparrow\downarrow\downarrow\rangle-
|\downarrow\uparrow\uparrow\rangle+|\downarrow\downarrow\downarrow\rangle)/2$%
, $|e_{2}\rangle=(-|\uparrow\uparrow\downarrow\rangle-|\uparrow\downarrow%
\downarrow\rangle+
|\downarrow\uparrow\uparrow\rangle+|\downarrow\downarrow\uparrow\rangle)/2$
and $|e_{3}\rangle=(-|\uparrow\downarrow\uparrow\rangle-|\uparrow\downarrow
\downarrow\rangle+|\downarrow\uparrow\uparrow\rangle
+|\downarrow\uparrow\downarrow\rangle)/2$. Under a global Hadamard gate, $%
|0\rangle$ turns to be $|\uparrow\rangle^{\otimes 3}$ and each sub-state of $%
|1\rangle$ is an entanglement of two spin-up and one spin-down. We may
accomplish the readout by detecting the spin-dependent spontaneously
emitting photons.

\bibitem{LD} D. Leibfried, R. Blatt, C. Monroe, and D. Wineland, Rev. Mod.
Phys. \textbf{75}, 281 (2003).

\bibitem{blatt} T. Monz, P. Schindler, J. T. Barreiro, M. Chwalla, D. Nigg,
W. A. Coish, M. Harlander, W. H\"{a}nsel, M. Hennrich, and R. Blatt, Phys.
Rev. Lett. \textbf{106}, 130506 (2011).

\bibitem{blatt2} B. P. Lanyon, C. Hempel, D. Nigg, M. M\"{u}ller, R.
Gerritsma, F. Z\"{a}hringer, P. Schindler, J. T. Barreiro, M. Rambach, G.
Kirchmair, M. Hennrich, P. Zoller, R. Blatt, and C. F. Roos, Science \textbf{%
334}, 57 (2011); D. P. DiVincenzo, Science \textbf{334}, 50 (2011).

\bibitem{riebe} M. Riebe, K. Kim, P. Schindler, T. Monz, P. O. Schmidt, T.
K. K\"{o}rber, W. H\"{a}nsel, H. H\"{a}ffner, C. F. Roos, and R. Blatt,
Phys. Rev. Lett. \textbf{97}, 220407 (2006).

\bibitem{geometric} M. V. Berry, Proc. Roy. Soc. Lond, Ser. A \textbf{392},
45 (1984); F. Wilczek and A. Zee, Phys. Rev. Lett. \textbf{52}, 2111 (1984);
Y. Aharonov and J. Anandan, Phys. Rev. Lett. \textbf{58}, 1593 (1987).

\bibitem{Erik} E. Sj\"{o}qvist, A. K. Pati, A. Ekert, J. S. Anandan, M.
Ericsson, D. K. L. Oi, and V. Vedral, Phys. Rev. Lett. \textbf{85}, 2845
(2000).

\bibitem{Wineland2} D. J. Wineland and D. Leibfried, Laser Phys. Lett.
\textbf{8}, 175 (2011).
\end{thebibliography}
\end{document}